\documentclass[a4paper]{iopart}

\usepackage{iopams}
\usepackage{setstack}
\usepackage{cite}
\usepackage{subeqn}
\usepackage{amsthm}
\usepackage{graphicx}

\newcommand{\Rii}{$R_{\mathrm{II}}$ }
\newcommand{\ol}{\overline}

\newtheorem{theorem}{Theorem}
\theoremstyle{remark}
\newtheorem{remark}{Remark}

\let\eqref\eref

\begin{document}
\title{A finite Toda representation of the box-ball system with box capacity}
\author{Kazuki Maeda}
\address{Department of Applied Mathematics and Physics, 
  Graduate School of Informatics, Kyoto University, Kyoto 606-8501, Japan}

\begin{abstract}
  A connection between the finite ultradiscrete Toda lattice
  and the box-ball system is extended to the case
  where each box has own capacity and a carrier has a capacity 
  parameter depending on time.
  In order to consider this connection, new carrier rules
  ``size limit for solitons'' and ``recovery of balls'', and a concept
  ``expansion map'' are introduced.
  A particular solution to the extended system 
  of a special case is also presented.
\end{abstract}

\ead{kmaeda@amp.i.kyoto-u.ac.jp}


\submitto{\it J. Phys. A: Math. Theor.}

\section{Introduction}
The box-ball system (BBS), proposed by Takahashi and Satsuma 
\cite{takahashi1990sca}, is one of the most important cellular automata
obtained from discrete integrable systems through a limiting procedure 
called ultradiscretization \cite{tokihiro1996fse}.
It is well known that the time evolution of the original BBS is determined
by the ultradiscrete KdV (u-KdV) equation:
\begin{equation}\label{eq:uKdV}
  U^{(t+1)}_n=\min\left(1-U^{(t)}_n, \sum_{j=-\infty}^{n-1} (U^{(t)}_j-U^{(t+1)}_j)\right),\quad n, t \in \mathbb Z,
\end{equation}
where $U^{(t)}_n \in \{0, 1\}$ denotes the number of balls in the $n$th box
at time $t$.
The equation \eqref{eq:uKdV} is an ultradiscrete analogue of 
the discrete KdV (d-KdV) equation, 
and consequently the original BBS has ultradiscrete soliton solutions.
It is also known that the nonautonomous discrete 
KP (nd-KP) equation with 2-reduction condition
yields a time evolution equation of the BBS with 
three extensions \cite{hatayama2001taa}:
box capacity \cite{takahashi1991casEn}, 
carrier capacity \cite{takahashi1997bbs}, 
and kind of balls \cite{tokihiro1999pos}.

\begin{figure}[htbp]
  \begin{center}
    \includegraphics{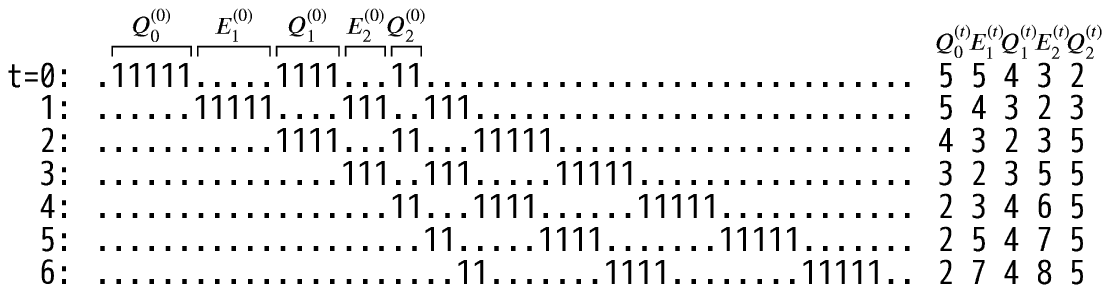}
    \caption{Example of the connection between the finite u-Toda lattice 
    and the original BBS. The variables $Q^{(t)}_n$ and $E^{(t)}_n$ denote
    the size of the $n$th soliton and of 
    the $n$th empty block at time $t$, respectively.}    
    \label{fig:exampleOfu-Toda}
  \end{center}
\end{figure}

In this paper, we consider another type of time evolution equations
of the BBS first presented by Nagai \etal\ \cite{nagai1999sca}.
They discovered that the ultradiscrete Toda (u-Toda)
equation {\em on a (non-periodic) finite lattice}
\begin{subequations}\label{eq:uToda}
  \begin{eqnarray}
    Q^{(t+1)}_n=\min\left(E^{(t)}_{n+1}, \sum_{j=0}^n Q^{(t)}_j-\sum_{j=0}^{n-1}Q^{(t+1)}_j\right),\\
    E^{(t+1)}_n=E^{(t)}_n-Q^{(t+1)}_{n-1}+Q^{(t)}_n,\\
    E^{(t)}_0=E^{(t)}_N=+\infty,
  \end{eqnarray}
\end{subequations}
which we refer to as the finite u-Toda lattice simply,
determines the time evolution of the original BBS.
In this equation, the dependent variables $Q^{(t)}_n$ and $E^{(t)}_n$
denote the size of the $n$th soliton at time $t$ and the size of the $n$th
empty block at time $t$, respectively.
\Fref{fig:exampleOfu-Toda} shows an example of the connection between
the finite u-Toda lattice and the BBS.
We call this representation {\em finite Toda representation} of the BBS.

The correspondence between the u-KdV equation and the finite u-Toda equation 
via the BBS is similar to the Euler-Lagrange correspondence 
of cellular automaton \cite{matsukidaira2003elc}.
This terminology comes from hydrodynamics;
the dependent variables of the Euler representation denote
the number of particles at each point and the ones of the Lagrange representation
denote the position of each particle.
According to these definitions, we use the following terms in this paper:
\begin{itemize}
\item Euler representation of BBS: the equation of the BBS with the variables
  which denote the number of balls in each box.
\item Lagrange representation of BBS: the equation of the BBS with the 
  variables which denote the start position of each soliton and each empty block.
\item Finite Toda representation of BBS: the equation of the BBS with the
  variables which denote the size of each soliton and each empty block.
\end{itemize}
The u-KdV equation \eqref{eq:uKdV} and the finite u-Toda lattice 
\eqref{eq:uToda} gives the Euler representation and the Lagrange representation 
of the original BBS, respectively.
Additionally, if we know the start position of the first soliton,
we can calculate the start positions of all solitons and empty blocks from 
the values of the variables of the finite Toda representation.
In other words, the finite Toda representation and the Lagrange representation
can be transformed to each other.

It has not been clarified by now why these two different ultradiscrete equations
with different type boundary conditions describe the same original BBS.
Moreover, the finite Toda representation of the extended BBSs
are not studied sufficiently.
Among the three extensions for the BBS, two types of the finite Toda representations 
have been clarified:
Tokihiro \etal \cite{tokihiro1999pos} discussed
the case of with several kinds of balls using the finite 
ultradiscrete hungry Toda lattice;
Tsujimoto and the author \cite{maeda2010bbs} showed that 
the finite nonautonomous ultradiscrete Toda (nu-Toda) lattice
determines the time evolution of the BBS with a carrier.
The main purpose of this paper is to discuss
the remaining case;
we derive a finite Toda representation of the BBS with box capacity.
For this purpose, we use a map from a state of the BBS
to a binary sequence, which we call ``expansion map''.
By using the expansion map, we can define the size of solitons in the 
BBS with box capacity for all time $t$, especially for interacting solitons.
We also consider the finite Toda representation of the BBS with 
both box capacity and carrier capacity using new carrier rules 
``size limit for solitons'' and ``recovery of balls''.
Furthermore, we present a particular solution for the fixed box capacity case.

The outline of this paper is the following.
In \sref{sec:euler-repr-bbs}, we recall the derivation of 
the Euler representation of the BBS with a carrier from the 2-reduced 
nd-KP equation.
In addition, we introduce new carrier rules ``size limit for solitons'' 
and ``recovery of balls''
which are used to derive the finite Toda representation of the BBS with 
a carrier \cite{maeda2010bbs} and play an important role 
in \sref{sec:extens-finite-toda}.
In \sref{sec:finite-toda-repr}, we recall the finite Toda representation
of the original BBS and give some remarks.
In \sref{sec:extens-finite-toda}, we discuss the finite Toda representation
of the BBS with variable box capacity.
First we discuss the case of variable box capacity $\Delta_n$ and 
no restricted carrier capacity.
After that, we discuss the case of both box capacity and carrier capacity are
variable.
In \sref{sec:part-solut-spec}, we give a particular solution to
the finite Toda representation of the BBS with fixed box capacity.
In \sref{sec:concluding-remarks}, we give concluding remarks.

\section{Euler representation of the BBS with a carrier}
\label{sec:euler-repr-bbs}

The nd-KP equation is given by \cite{willox1997dbd}
\begin{equation}\label{eq:nd-KP}
  \fl
  (a_n-b_t)f^{k, t+1}_{n+1}f^{k+1, t}_n
  +(b_t-c_k)f^{k, t}_{n+1}f^{k+1, t+1}_n
  +(c_k-a_n)f^{k, t+1}_n f^{k+1, t}_{n+1}=0,\quad 
  k, n, t \in \mathbb Z.
\end{equation}
It is shown that the $N$-soliton solution to the nd-KP equation 
\eref{eq:nd-KP} is presented by
\begin{eqnarray*}
  \fl f^{k, t}_n=1+\sum_{\substack{J \subset\{0, 1, \dots, N-1\} \cr J \ne \emptyset}}\left(\prod_{\substack{i, j \in J \cr i \ne j}}w_{i, j}\prod_{i \in J} h^{k, t}_{i, n}\right),\\
  \fl h^{k, t}_{i, n}:=\xi_i\prod_{j=0}^{n-1}\frac{a_j-p_i}{a_j-q_i}\prod_{j=0}^{t-1}\frac{b_j-p_i}{b_j-q_i}\prod_{j=0}^{k-1}\frac{c_j-p_i}{c_j-q_i},\quad
  w_{i, j}:=\frac{(p_i-p_j)(q_i-q_j)}{(p_i-q_j)(q_i-p_j)},
\end{eqnarray*}
where $\xi_i$, $p_i$ and $q_i$, $i=0, 1, \dots, N-1$, are some constants.
Now we impose the 2-reduction condition with respect to the variable $k$,
that is $f^{k+2, t}_n=f^{k, t}_n$ and $c_{k+2}=c_k$ for all $k \in \mathbb Z$,
and set $a_n=1+\delta_n$, $b_t=-\mu_t$, $c_0=1$ and $c_1=0$.
Then the nd-KP equation \eref{eq:nd-KP} reduces to the forms
\begin{subequations}\label{eq:2-reduced nd-KP}
  \begin{eqnarray}
    (1+\delta_n+\mu_t)f^{0, t+1}_{n+1}f^{1, t}_n
    =(1+\mu_t)f^{0, t}_{n+1}f^{1, t+1}_n
    +\delta_n f^{0, t+1}_n f^{1, t}_{n+1},\\
    (1+\delta_n+\mu_t)f^{0, t}_n f^{1, t+1}_{n+1}
    =(1+\delta_n)f^{0, t}_{n+1}f^{1, t+1}_n
    +\mu_t f^{0, t+1}_n f^{1, t}_{n+1},
  \end{eqnarray}
\end{subequations}
and an $N$-soliton solution to the reduced equations is given by
\begin{subequations}\label{eq:2rnd-KP soliton}
  \begin{eqnarray}
    f^{k, t}_n=1+\sum_{\substack{J \subset\{0, 1, \dots, N-1\} \cr J \ne \emptyset}}\left(\prod_{\substack{i, j \in J \cr i \ne j}}w_{i, j}\prod_{i \in J} h^{k, t}_{i, n}\right),\quad k=0, 1,\\
    h^{0, t}_{i, n}:=\xi_i\prod_{j=0}^{n-1}\frac{1+\delta_j-p_i}{p_i+\delta_j}\prod_{j=0}^{t-1}\frac{p_i+\mu_j}{1+\mu_j-p_i},\quad
    h^{1, t}_{i, n}:=\frac{1-p_i}{p_i}h^{0, t}_{i, n},\\
    w_{i, j}:=\left(\frac{p_i-p_j}{1-p_i-p_j}\right)^2.
  \end{eqnarray}
\end{subequations}

Let us define the dependent variables as
\begin{equation}\label{eq:2rnd-KP tau}
  \fl u^{(t)}_n=\frac{f^{0, t+1}_{n+1}f^{1, t+1}_n}{f^{0, t+1}_n f^{1, t+1}_{n+1}},\quad
  \ol u^{(t)}_n=(1+\delta_n+\mu_t)\frac{f^{0, t}_n f^{0, t+1}_{n+1}}{f^{0, t}_{n+1}f^{0, t+1}_n},\quad
  \ol z^{(t)}_n=\frac{f^{0, t}_{n}f^{1, t+1}_{n}}{f^{0, t+1}_{n}f^{1, t}_{n}}.
\end{equation}
Then the 2-reduced nd-KP equation \eref{eq:2-reduced nd-KP} yields
the equations
\begin{subequations}\label{eq:2rnd-KP}
  \begin{eqnarray}
    \ol u^{(t+1)}_n=\delta_n\frac{1}{u^{(t)}_n}+(1+\mu_{t+1})\ol z^{(t+1)}_{n},\\
    \ol z^{(t+1)}_n=\frac{(1+\delta_{n})\ol z^{(t+1)}_{n-1}u^{(t)}_{n-1}+\mu_{t+1}}{\ol u^{(t+1)}_{n-1}},
  \end{eqnarray}
  and the identity
  \begin{equation}
    u^{(t+1)}_n=u^{(t)}_n\frac{\ol z^{(t+1)}_{n}}{\ol z^{(t+1)}_{n+1}}
  \end{equation}
  holds.
\end{subequations}
For positivity, we choose the parameters as $0 \le \delta_n \le 1$ and
$0 \le \mu_t \le 1$ for all $n, t \in \mathbb Z$.
When the values of the dependent variables
are all positive for all $n, t \in \mathbb Z$,
we can ultradiscretize the equations
\eref{eq:2rnd-KP}: putting
$u^{(t)}_n=\rme^{-U^{(t)}_n/\epsilon}$, 
$\ol u^{(t)}_n=\rme^{-\ol U^{(t)}_n/\epsilon}$, 
$\ol z^{(t)}_n=\rme^{-\ol Z^{(t)}_n/\epsilon}$,
$\delta_n=\rme^{-\Delta_n/\epsilon}$,
$\mu_t=\rme^{-M_t/\epsilon}$ into \eqref{eq:2rnd-KP}
and taking a limit $\epsilon \to +0$, we obtain the ultradiscrete system
\begin{subequations}\label{eq:2rnuKp}
  \begin{eqnarray}
    \ol U^{(t+1)}_n=\min(\Delta_n-U^{(t)}_n, \ol Z^{(t+1)}_{n}),\label{eq:2rnu-KP 1}\\
    \ol Z^{(t+1)}_n=\min(\ol Z^{(t+1)}_{n-1}+U^{(t)}_{n-1}, M_{t+1})-\ol U^{(t+1)}_{n-1},\label{eq:2rnu-KP 2}\\
    U^{(t+1)}_n=U^{(t)}_n+\ol Z^{(t+1)}_{n}-\ol Z^{(t+1)}_{n+1}, \label{eq:2rnu-KP 3}
  \end{eqnarray}
\end{subequations}
where $\Delta_n, M_t \ge 0$ for all $n, t \in \mathbb Z$.
Note that we have used the fundamental formula for the ultradiscretization:
\begin{equation*}
  \lim_{\epsilon \to +0} -\epsilon\log(\rme^{-A/\epsilon}+\rme^{-B/\epsilon})
  =\min(A, B).
\end{equation*}
An $N$-soliton solution to the ultradiscrete system \eqref{eq:2rnuKp} is 
obtained as follows.
Let us take the constants $p_i$, $i=0, 1, \dots, N-1$, to satisfy 
the condition $0<p_i<1$.
Putting $f^{k, t}_n=\rme^{-F^{k, t}_n/\epsilon}$,
$h^{k, t}_{i, n}=\rme^{-H^{k, t}_{i, n}/\epsilon}$,
$p_i=\rme^{-P_i/\epsilon}$,
$\xi_i=\rme^{-\Xi_i/\epsilon}$,
$w_{i, j}=\rme^{-W_{i, j}/\epsilon}$ into 
\eqref{eq:2rnd-KP soliton} and \eqref{eq:2rnd-KP tau},
and taking a limit $\epsilon \to +0$, we obtain
\begin{eqnarray*}
  U^{(t)}_n=F^{0, t+1}_{n+1}-F^{0, t+1}_n+F^{1, t+1}_n-F^{1, t+1}_{n+1},\\
  \ol U^{(t)}_n=F^{0, t}_n-F^{0, t}_{n+1}+F^{0, t+1}_{n+1}-F^{0, t+1}_n,\\
  \ol Z^{(t)}_n=F^{0, t}_n-F^{0, t+1}_n+F^{1, t+1}_{n}-F^{1, t}_{n},\\
  F^{k, t}_n=\min\left(0, \min_{\substack{J \subset\{0, 1, \dots, N-1\} \cr J \ne \emptyset}}\left(\sum_{\substack{i, j \in J\cr i \ne j}}W_{i, j}+\sum_{i \in J}H^{k, t}_{i, n}\right)\right),\quad k=0, 1,\\
  H^{0, t}_{i, n}=\Xi_i-\sum_{j=0}^{n-1}\min(P_i, \Delta_j)+\sum_{j=0}^{t-1}\min(P_i, M_j),\quad
  H^{1, t}_{i, n}=H^{0, t}_{i, n}-P_i,\\
  W_{i, j}=2\min(P_i, P_j),
\end{eqnarray*}
and $P_i \ge 0$, $i=0, 1, \dots, N-1$.

\begin{figure}[htbp]
  \begin{center}
    \includegraphics{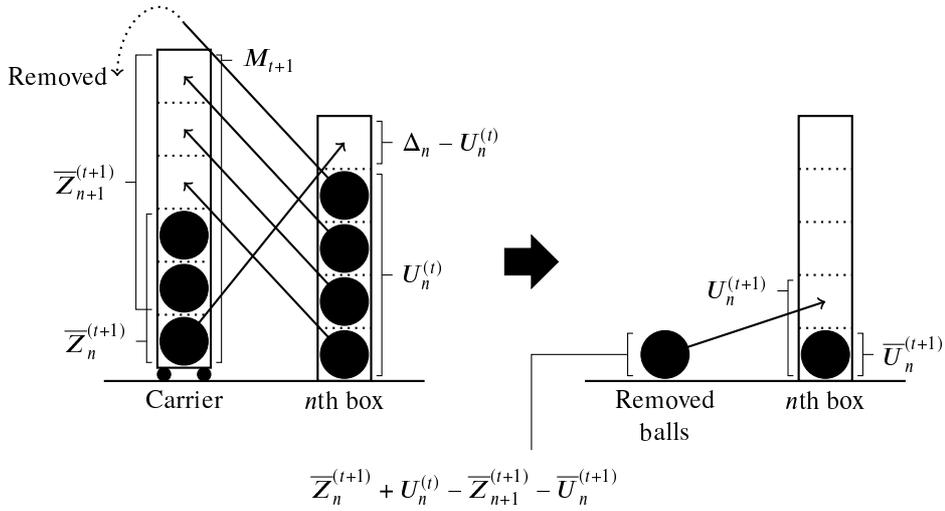}
    \caption{Illustration of the time evolution rule of the BBS with a carrier.
      The left figure illustrates the size limit process \eqref{eq:2rnu-KP 1} 
      and \eqref{eq:2rnu-KP 2}, and the right one illustrates the 
      recovery process \eqref{eq:2rnu-KP 3}.
    }
    \label{fig:IllustOfTheRule}
  \end{center}
\end{figure}

Let us introduce the time evolution rule of 
the BBS with the $n$th box capacity $\Delta_n$ and the 
carrier capacity $M_{t+1}$ from time $t$ to $t+1$.
We consider the time evolution rule from time $t$ to $t+1$
as the composition of {\em size limit process} and {\em recovery process}.
\begin{enumerate}
\item Size limit process: 
  the carrier of balls moves from left ($n=-\infty$)
  to right ($n=+\infty$).
  When the carrier passes each box, the carrier gets all balls in the box;
  and if the number of balls exceeds the carrier capacity $M_{t+1}$,
  the excess balls are removed from the system.
  At the same time, the carrier puts the balls which the carrier holds
  into the box as many as possible.
\item Recovery process: after the size limit process, all the removed balls
  are recovered to the boxes in which the balls were.
\end{enumerate}
\Fref{fig:IllustOfTheRule} illustrates these rules.

Suppose that the dependent variables $U^{(t)}_n$, 
$\ol U^{(t+1)}_n$ and $\ol Z^{(t+1)}_n$ denote the following quantities:
\begin{itemize}
\item $U^{(t)}_n \in \{0, 1, \dots, \Delta_n\}$: 
  the number of balls in the $n$th box at time $t$;
\item $\ol U^{(t+1)}_n \in \{0, 1, \dots, \Delta_n\}$: 
  the number of balls in the $n$th box after the 
  size limit process from time $t$ to $t+1$;
\item $\ol Z^{(t+1)}_n \in \{0, 1, \dots, M_{t+1}\}$: 
  the number of balls in the carrier arriving at the $n$th box
  in the size limit process from time $t$ to $t+1$.
\end{itemize}
Then the equations \eqref{eq:2rnuKp} give the time evolution rule:
the equations \eqref{eq:2rnu-KP 1} and \eqref{eq:2rnu-KP 2} define
the size limit process 
\begin{equation*}
  \{U^{(t)}_n\}_{n=-\infty}^{+\infty}\mapsto (\{\ol U^{(t+1)}_n\}_{n=-\infty}^{+\infty}, \{\ol Z^{(t+1)}_n\}_{n=-\infty}^{+\infty})
\end{equation*}
and, since
\begin{eqnarray*}
  \fl&\phantom{{}={}}\ol U^{(t+1)}_{n}+\left((\ol Z^{(t+1)}_n+U^{(t)}_n)-\min(\ol Z^{(t+1)}_n+U^{(t)}_n, M_{t+1})\right)\\
  &=\ol U^{(t+1)}_n+\ol Z^{(t+1)}_{n}+U^{(t)}_n-\ol Z^{(t+1)}_{n+1}-\ol U^{(t+1)}_n\\
  &=U^{(t)}_n+\ol Z^{(t+1)}_n-\ol Z^{(t+1)}_{n+1}
\end{eqnarray*}
gives the number of removed balls by the size limit at the $n$th box,
the equation \eqref{eq:2rnu-KP 3} defines the recovery process
\begin{equation*}
  (\{\ol U^{(t+1)}_n\}_{n=-\infty}^{+\infty}, \{\ol Z^{(t+1)}_n\}_{n=-\infty}^{+\infty}) \mapsto \{U^{(t+1)}_n\}_{n=-\infty}^{+\infty}.
\end{equation*}

\begin{figure}[htbp]
  \begin{center}
    \includegraphics{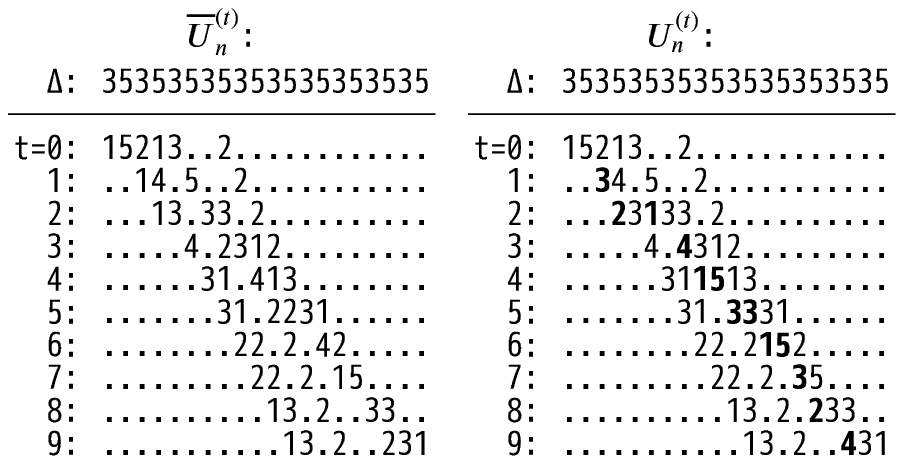}
    \caption{Example of a 3-soliton solution to 
      the time evolution equation (\ref{eq:2rnuKp}).
      The leftmost box is the $0$th box in the both figures.
      The carrier capacity $M_t=6$ for all $t \ge 1$.
      Each number denotes the number of balls in a box
      and `\texttt{.}' denotes an empty box.
      In the right figure, boxes containing recovered balls
      are shown in boldface (compare to the left figure).}
    \label{fig:exampleOf2rnu-KP}
  \end{center}
\end{figure}

\Fref{fig:exampleOf2rnu-KP} shows an example of a 3-soliton solution
to the time evolution equation \eqref{eq:2rnuKp}.
The parameters are chosen as
\begin{equation*}
  \Delta_n=
  \cases{
    3 & \text{if $n$ is even},\\
    5 & \text{if $n$ is odd},    
  }\quad
  M_t=
  \cases{
    +\infty & \text{if $t \le 0$},\\
    6 & \text{if $t>0$}.
  }
\end{equation*}

\begin{remark}
  Eliminating the variable $\ol U^{(t+1)}_n$ from the equations
  \eqref{eq:2rnu-KP 1} and \eqref{eq:2rnu-KP 2}, we have the relation
  \begin{equation}\label{eq:pre-umKdV}
    -\ol Z^{(t+1)}_{n+1}=\min(\Delta_n-U^{(t)}_n, \ol Z^{(t+1)}_n)-\min(\ol Z^{(t+1)}_{n}+U^{(t)}_n, M_{t+1}).
  \end{equation}
  From the equation \eqref{eq:2rnu-KP 3}, the relation
  \begin{equation}\label{eq:z-infinite-sum}
    \ol Z^{(t+1)}_n=\sum_{j=-\infty}^{n-1}(U^{(t)}_j-U^{(t+1)}_j)
  \end{equation}
  also holds.
  Substituting \eqref{eq:z-infinite-sum} into \eqref{eq:pre-umKdV}, we obtain
  the equation
  \begin{equation}\label{eq:umKdV}
    \eqalign{
      U^{(t+1)}_n&=\min\left(\Delta_n-U^{(t)}_n, \sum_{j=-\infty}^{n-1} (U^{(t)}_j-U^{(t+1)}_j)\right)\\
      &\qquad+\max\left(0, \sum_{j=-\infty}^n U^{(t)}_j-\sum_{j=-\infty}^{n-1}U^{(t+1)}_j-M_{t+1}\right),
    }
  \end{equation}
  where we have used the formula
  \begin{equation*}
    -\min(-A, -B)=\max(A, B).
  \end{equation*}
  The equation \eqref{eq:umKdV} has the same form as of 
  the time evolution equation of 
  the ``BBS with a carrier'' presented by
  Takahashi and Matsukidaira \cite{takahashi1997bbs}.

  If we choose $M_{t+1}=+\infty$, then \eqref{eq:umKdV} yields
  \begin{equation}\label{eq:nu-KdV}
    U^{(t+1)}_n=\min\left(\Delta_n-U^{(t)}_n, \sum_{j=-\infty}^{n-1} (U^{(t)}_j-U^{(t+1)}_j)\right),
  \end{equation}
  which is the nonautonomous u-KdV equation.
  In addition, \eqref{eq:2rnu-KP 2} yields
  \begin{equation*}
    \ol Z^{(t+1)}_n
    =\ol Z^{(t+1)}_{n-1}+U^{(t)}_{n-1}-\ol U^{(t+1)}_{n-1},
  \end{equation*}
  and comparing this relation with \eqref{eq:2rnu-KP 3},
  we have the relation $U^{(t+1)}_n=\ol U^{(t+1)}_n$.
\end{remark}

\section{Finite Toda representation of the original BBS}
\label{sec:finite-toda-repr}

We recall the relation between 
the finite u-Toda lattice and the original BBS.
The bilinear form of the discrete Toda lattice is given by
\begin{equation}\label{eq:d-Toda-bilinear}
  \tau^{(t-1)}_n\tau^{(t+1)}_n=\tau^{(t-1)}_{n+1}\tau^{(t+1)}_{n-1}+\tau^{(t)}_n\tau^{(t)}_n,\quad n, t \in \mathbb Z.
\end{equation}
Let us introduce the dependent variables
\begin{equation*}
  q^{(t)}_n=\frac{\tau^{(t)}_n\tau^{(t+1)}_{n+1}}{\tau^{(t)}_{n+1}\tau^{(t+1)}_n},\quad
  e^{(t)}_n=\frac{\tau^{(t)}_{n+1}\tau^{(t+1)}_{n-1}}{\tau^{(t)}_n\tau^{(t+1)}_n},\quad
  d^{(t)}_n=\frac{\tau^{(t)}_n\tau^{(t)}_{n+1}}{\tau^{(t-1)}_{n+1}\tau^{(t+1)}_n}.
\end{equation*}
Then \eqref{eq:d-Toda-bilinear} yields the equation
\begin{subequations}\label{eq:dqd}
  \begin{equation}
    q^{(t+1)}_n=e^{(t)}_{n+1}+d^{(t+1)}_n,
  \end{equation}
  and the identities
  \begin{equation}
    e^{(t+1)}_n=e^{(t)}_n\frac{q^{(t)}_n}{q^{(t+1)}_{n-1}},\quad
    d^{(t+1)}_n=d^{(t+1)}_{n-1}\frac{q^{(t)}_n}{q^{(t+1)}_{n-1}}
  \end{equation}
\end{subequations}
hold.
Putting $q^{(t)}_n=\rme^{-Q^{(t)}_n/\epsilon}$,
$e^{(t)}_n=\rme^{-E^{(t)}_n/\epsilon}$,
$d^{(t)}_n=\rme^{-D^{(t)}_n/\epsilon}$,
and taking a limit $\epsilon \to +0$,
we obtain the u-Toda lattice
\begin{subequations}\label{eq:finite-u-Toda}
  \begin{eqnarray}
    Q^{(t+1)}_n=\min(E^{(t)}_{n+1}, D^{(t+1)}_n),\\
    E^{(t+1)}_n=E^{(t)}_n-Q^{(t+1)}_{n-1}+Q^{(t)}_n,\\
    D^{(t+1)}_n=D^{(t+1)}_{n-1}-Q^{(t+1)}_{n-1}+Q^{(t)}_n\label{eq:finite-u-Toda-D}.
  \end{eqnarray}
  Furthermore, we impose the terminating condition for
  discussing the finite Toda representation:
  \begin{equation}\label{eq:finite-u-Toda-BC}
    E^{(t)}_0=E^{(t)}_N=+\infty,\quad D^{(t+1)}_0=Q^{(t)}_0,
  \end{equation}
  where $N$ is a positive integer, which denotes the number of solitons
  in the original BBS.
\end{subequations}

Let the variables $Q^{(t)}_n$, $E^{(t)}_n$ and $D^{(t+1)}_n$, respectively,
denote the following quantities of the original BBS:
\begin{itemize}
\item $Q^{(t)}_n$: the size of the $n$th soliton at time $t$
  ($n=0, 1, \dots, N-1$);
\item $E^{(t)}_n$: the size of the $n$th empty block,
  namely, the distance between the ($n-1$)th soliton and the $n$th one
  at time $t$ ($n=1, 2, \dots, N-1$);
\item $D^{(t+1)}_n$: the number of balls in the carrier
  after getting $Q^{(t)}_n$ balls ($n=0, 1, \dots, N-1$).
\end{itemize}
Then the next theorem gives a fundamental result on the connection
between the finite u-Toda lattice and the BBS.

\begin{theorem}[Nagai \etal \cite{nagai1999sca}]\label{th:u-TodaAndBBS}
  The finite u-Toda lattice \eqref{eq:finite-u-Toda} determines 
  the time evolution of the original BBS.
\end{theorem}

\begin{remark}\label{rem:negative-problem}
  Conventionally, the d-Toda lattice (qd-type) has been written in the form
  \begin{equation*}
    q^{(t+1)}_n+e^{(t+1)}_n=q^{(t)}_n+e^{(t)}_{n+1},\quad
    q^{(t+1)}_{n-1}e^{(t+1)}_n=q^{(t)}_n e^{(t)}_n,
  \end{equation*}
  or
  \begin{subequations}\label{eq:qd}
    \begin{eqnarray}
      q^{(t+1)}_n=q^{(t)}_n-e^{(t+1)}_n+e^{(t)}_{n+1},\label{eq:qda}\\
      e^{(t+1)}_n=e^{(t)}_n\frac{q^{(t)}_n}{q^{(t+1)}_{n-1}},\label{eq:qdm}
    \end{eqnarray}
  \end{subequations}
  which we cannot ultradiscretize directly due to ``negative problem''.
  The equations \eqref{eq:qd} are called Rutishauser's qd algorithm
  in numerical algorithms \cite{rutishauser1990lon}.
  On the other hand, the equations \eqref{eq:dqd} are called dqd algorithm.
  The dqd algorithm is the subtraction-free form of the qd algorithm
  and computes matrix eigenvalues or singular values more accurately
  than the qd algorithm.

  Suppose the finite lattice condition $e^{(t)}_0=e^{(t)}_N=0$.
  Nagai \etal \cite{nagai1999sca} rewrote \eqref{eq:qda} using \eqref{eq:qdm} as
  \begin{eqnarray*}
    q^{(t+1)}_n
    &=e^{(t)}_{n+1}+q^{(t)}_n-e^{(t+1)}_n\\
    &=e^{(t)}_{n+1}+\frac{q^{(t)}_n}{q^{(t+1)}_{n-1}}(q^{(t+1)}_{n-1}-e^{(t)}_n)\\
    &=e^{(t)}_{n+1}+\frac{q^{(t)}_n}{q^{(t+1)}_{n-1}}(q^{(t)}_{n-1}-e^{(t+1)}_{n-1})\\
    &=\dots\\
    &=e^{(t)}_{n+1}+\frac{\prod_{j=0}^n q^{(t)}_j}{\prod_{j=0}^{n-1}q^{(t+1)}_j}.
  \end{eqnarray*}
  Then they could ultradiscretize the finite Toda lattice:
  \begin{subequations}\label{eq:u-Toda}
  \begin{eqnarray}
    Q^{(t+1)}_n=\min\left(E^{(t)}_{n+1}, \sum_{j=0}^n Q^{(t)}_j-\sum_{j=0}^{n-1}Q^{(t+1)}_j\right),\\
    E^{(t+1)}_n=E^{(t)}_n-Q^{(t+1)}_{n-1}+Q^{(t)}_n,\\
    E^{(t)}_0=E^{(t)}_N=+\infty.
  \end{eqnarray}
  \end{subequations}
  On the other hand, by introducing an auxiliary variable 
  \begin{equation*}
    d^{(t+1)}_n:=q^{(t)}_n-e^{(t+1)}_n=q^{(t+1)}_n-e^{(t)}_{n+1},
  \end{equation*}
  we can ultradiscretize the finite d-Toda lattice directly without the negative
  problem and obtain 
  the finite u-Toda lattice of the dqd form \eqref{eq:finite-u-Toda}.
  From the viewpoint of the BBS, the variable $D^{(t+1)}_n$ denotes the number
  of balls in the carrier. Therefore, the finite u-Toda lattice of the dqd
  form \eqref{eq:finite-u-Toda} is important to consider the finite Toda
  representation of the BBS with a carrier.
\end{remark}

\begin{figure}[htbp]
  \begin{center}
    \includegraphics[scale=.825]{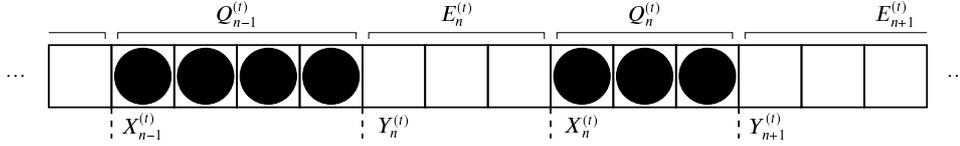}
    \caption{Lagrange representation and finite Toda representation of the BBS.}
    \label{fig:Lagrange-Toda}
  \end{center}
\end{figure}

\begin{remark}\label{rem:Lagrange}
  Here we remark on the Lagrange representation of the BBS,
  which is also a terminology from hydrodynamics as the Euler representation;
  the dependent variables of the Lagrange representation 
  denote the position of solitons.
  Let the variables $X^{(t)}_n$ and $Y^{(t)}_n$ denote
  the start position of the $n$th soliton and 
  the one of the $n$th empty block at time $t$, respectively 
  (see \Fref{fig:Lagrange-Toda}).
  Then the Lagrange representation of the BBS is given by 
  \cite{matsukidaira2005elcen}
  \begin{subequations}\label{eq:Lagrange}
    \begin{eqnarray}
      \fl X^{(t+1)}_n=Y^{(t)}_{n+1},\label{eq:Lagrange1}\\
      \fl Y^{(t+1)}_n=Y^{(t)}_n+\min\left(X^{(t)}_n-Y^{(t)}_n, \sum_{j=1}^n (Y^{(t)}_j-X^{(t)}_{j-1})-\sum_{j=1}^{n-1}(Y^{(t+1)}_j-X^{(t+1)}_{j-1})\right),\\
      \fl Y^{(t)}_0=-\infty,\quad X^{(t)}_N=+\infty.
    \end{eqnarray}
  \end{subequations}
  Relations between these variables and the variables of the finite u-Toda
  lattice \eqref{eq:u-Toda} are given by
  \begin{equation}\label{eq:LagrangeAndu-Toda}
    X^{(t)}_n=Y^{(t)}_n+E^{(t)}_n,\quad
    Y^{(t)}_n=X^{(t)}_{n-1}+Q^{(t)}_{n-1}.
  \end{equation}
  We can readily show that \eqref{eq:Lagrange} and \eqref{eq:LagrangeAndu-Toda}
  yield the finite u-Toda lattice \eqref{eq:u-Toda}.
  Conversely, we can calculate the values of 
  $\{X^{(t)}_n\}_{n=1}^{N-1}$ and $\{Y^{(t)}_n\}_{n=1}^{N}$ from 
  the values of $X^{(t)}_0$, $\{Q^{(t)}_n\}_{n=0}^{N-1}$
  and $\{E^{(t)}_n\}_{n=1}^{N-1}$ using
  the relations \eqref{eq:LagrangeAndu-Toda}.
  In other words, the finite u-Toda lattice \eqref{eq:u-Toda} and
  the equation $X^{(t+1)}_0=X^{(t)}_0+Q^{(t)}_0$, which is obtained
  from \eqref{eq:Lagrange1} and \eqref{eq:LagrangeAndu-Toda},
  uniquely determine the time evolution of the BBS.
\end{remark}

\section{Extension of the finite Toda representation to the case of variable box capacity $\Delta_n$ and variable carrier capacity $M_t$}
\label{sec:extens-finite-toda}

In previous studies, the finite Toda representation is considered only 
for the BBS with box capacity 1.
In this section, we extend the finite Toda representation
to the case in which each box has own capacity $\Delta_n$.
First we consider the case of carrier capacity $M_t=+\infty$.
The Euler representation of this case is given by \eqref{eq:nu-KdV}.

We first define the size of solitons and the one of
empty blocks for the BBS with variable box capacity $\Delta_n$ at any time $t$.
For this purpose, we refer to the work of Takahashi and 
Satsuma \cite{takahashi1991casEn}.
They analyzed the BBS with the fixed box capacity $\Delta$ using a map
from a state of box capacity $\Delta$ to a binary sequence.
We generalize this map for the case of variable box capacity $\Delta_n$.

Suppose that a state of the Euler representation \eqref{eq:nu-KdV}
$\{U^{(t)}_n\}_{n=-\infty}^{+\infty}$ such that
$U^{(t)}_n \in \{0, 1, \dots, \Delta_n\}$ is given.
We assume that, for simplicity, $U^{(t)}_n=0$ for $n<0$.
Let us define a map $\{U^{(t)}_n\}_{n=-\infty}^{+\infty} \mapsto
\{V^{(t)}_n\}_{n=-\infty}^{+\infty}$, where $V^{(t)}_n \in \{0, 1\}$,
as follows:
\begin{enumerate}
\renewcommand{\labelenumi}{(\arabic{enumi})}
\item $V^{(t)}_n=0$ for $n<0$.
\item Let $s_0=0$ and $s_n=\sum_{j=0}^{n-1}\Delta_j$ for $n=1, 2,\dots$. 
  From $n=0$ to $+\infty$, if $V^{(t)}_{s_n-1}=1$, then 
  \begin{eqnarray*}
    V^{(t)}_{s_n}=V^{(t)}_{s_n+1}=\dots=V^{(t)}_{s_n+U^{(t)}_n-1}=1,\\
    V^{(t)}_{s_n+U^{(t)}_n}=V^{(t)}_{s_n+U^{(t)}_n+1}=\dots=V^{(t)}_{s_n+\Delta_n-1}=0;
  \end{eqnarray*}
  otherwise, 
  \begin{eqnarray*}
    V^{(t)}_{s_n}=V^{(t)}_{s_n+1}=\dots=V^{(t)}_{s_n-U^{(t)}_n-1+\Delta_n}=0,\\
    V^{(t)}_{s_n-U^{(t)}_n+\Delta_n}=V^{(t)}_{s_n-U^{(t)}_n+\Delta_n+1}=\dots=V^{(t)}_{s_n+\Delta_n-1}=1.
  \end{eqnarray*}
  Note that the relation
  $U^{(t)}_n=\sum_{j=s_n}^{s_n+\Delta_n-1} V^{(t)}_j$ holds.
\end{enumerate}

We refer the $j$th number $V^{(t)}_j$ in the binary sequence 
as the $j$th {\em segment}.
By using this map, the $n$th box is expanded to the block composed from the 
$s_n$th
to $(s_n+\Delta_n-1)$th segments.
We call this map {\em expansion map} from a state of the BBS with
variable box capacity $\Delta_n$ to a binary sequence.

\begin{figure}[htbp]
  \begin{center}
    \includegraphics{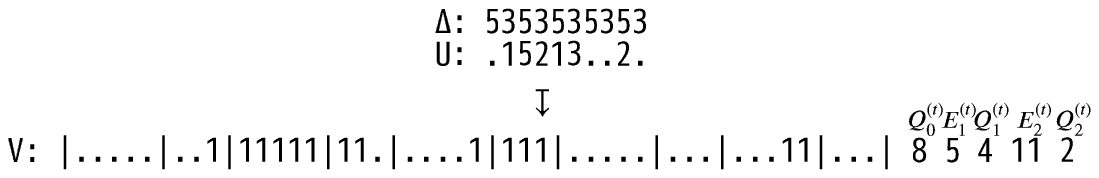}
    \caption{Example of the expansion map from a state of 
      the BBS with variable box capacity $\Delta_n$ to a binary sequence.
      In the binary sequence, a block 
      between two separators `{\tt |}' corresponds to an original box.}
    \label{fig:exampleOfMap}
    
    \vspace{1em}

    \includegraphics[scale=1.1]{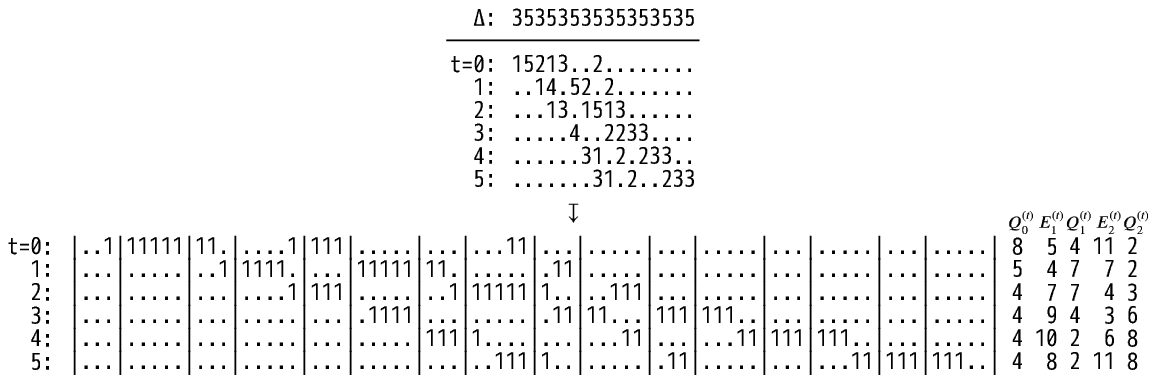}
    \caption{Example of the expansion map for the BBS with
      variable box capacity $\Delta_n$. The carrier capacity parameter is chosen as
      $M_t=+\infty$ for all $t \in \mathbb Z$.}
    \label{fig:exampleOfMapEx}
  \end{center}
\end{figure}

Figures~\ref{fig:exampleOfMap} and \ref{fig:exampleOfMapEx} show
examples of the expansion map.
As shown in Figure~\ref{fig:exampleOfMapEx},
the expansion map enables us to define 
the size of the $n$th soliton $Q^{(t)}_n$ and
the one of the $n$th empty block $E^{(t)}_n$ 
for the BBS with box capacity $\Delta_n$ at any time $t$
in the same way as for the BBS with box capacity $1$.
Let $D^{(t+1)}_n$ denote the number of balls
which the carrier holds after getting $Q^{(t)}_n$ balls,
and $\Lambda^{(t)}_n$ denote the capacity of the box which 
contains the beginning (leftmost) segment of the $n$th empty block.
Then we arrive at the following theorem.
\begin{theorem}
  Let the variables $Q^{(t)}_n$, $E^{(t)}_n$ and $D^{(t+1)}_n$ denote
  the quantities of the BBS as explained in the previous section.
  Then the time evolution of the BBS with variable box capacity $\Lambda^{(t)}_n$
  is given by
  \begin{subequations}\label{eq:exToda}
    \begin{eqnarray}
      Q^{(t+1)}_n=\min\left(E^{(t)}_{n+1}-\max(0, \Lambda^{(t)}_{n+1}-D^{(t+1)}_n), D^{(t+1)}_n\right),\label{eq:exToda-1}\\
      \eqalign{
      E^{(t+1)}_n&=E^{(t)}_n-Q^{(t+1)}_{n-1}+Q^{(t)}_n\\
      &\qquad-\max(0, \Lambda^{(t)}_n-D^{(t+1)}_{n-1})+\max(0, \Lambda^{(t)}_{n+1}-D^{(t+1)}_n),}\label{eq:exToda-2}\\
      D^{(t+1)}_n=D^{(t+1)}_{n-1}-Q^{(t+1)}_{n-1}+Q^{(t)}_n,\label{eq:exToda-3}\\
      E^{(t)}_0=E^{(t)}_N=+\infty,\quad
      D^{(t+1)}_0=Q^{(t)}_0.
    \end{eqnarray}
  \end{subequations}
\end{theorem}

We note that, from \eqref{eq:exToda-1},
$Q^{(t+1)}_n \le D^{(t+1)}_n$ holds for all $n$ and $t$.
Since the size of the $n$th soliton $Q^{(t)}_n$ 
should be positive for all $n$ and $t$,
from \eqref{eq:exToda-3},
the inequality $D^{(t+1)}_n \ge 1$ holds for all $n$ and $t$.
Thus, all $\max(0, \Lambda^{(t)}_{n+1}-D^{(t+1)}_n)$ are equal to zero
when $\Lambda^{(t)}_n=1$ for all $n$ and $t$,
the case of the original BBS.
In this case, the equations \eqref{eq:exToda} 
reduce to the finite Toda representation of 
the original BBS \eqref{eq:finite-u-Toda}.
Hence we can say that the ultradiscrete system \eqref{eq:exToda} is
a generalization of the finite u-Toda lattice \eqref{eq:finite-u-Toda}.

\begin{proof}  
  Let us consider the general $\Lambda^{(t)}_n \ge 1$ case.
  As we mentioned above, \eqref{eq:exToda} has additional terms 
  $\max(0, \Lambda^{(t)}_{n+1}-D^{(t+1)}_n)$ which do not appear
  in the case of box capacity $1$ \eqref{eq:finite-u-Toda}.
  Hence, we shall investigate the role of the terms
  $\max(0, \Lambda^{(t)}_{n+1}-D^{(t+1)}_n)$.

  \begin{figure}[htbp]
    \begin{center}
      \includegraphics[scale=.7]{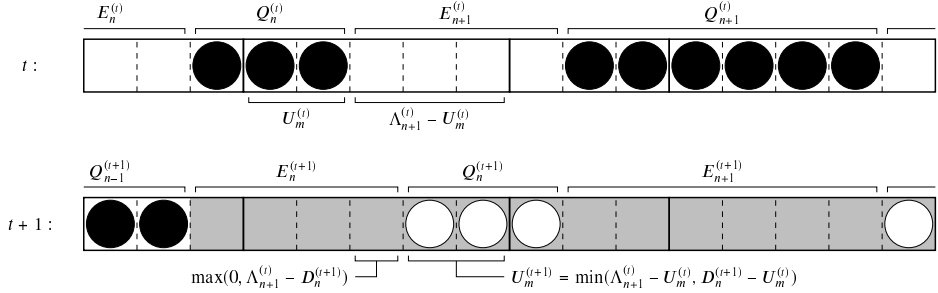}
      \caption{Illustration of the quantity $\max(0, \Lambda^{(t)}_{n+1}-D^{(t+1)}_n)$.
        We now need to determine the quantities on the area filled 
        with gray: $E^{(t+1)}_n$, $Q^{(t+1)}_n$, $E^{(t+1)}_{n+1}$, $\dots$, 
        $Q^{(t+1)}_{N-1}$.
        In this figure, $D^{(t+1)}_{n-1}-Q^{(t+1)}_{n-1}=1$ is assumed;
        the carrier has one ball just before getting $Q^{(t)}_n$ 
        balls.}
      \label{fig:boxcap-ex-illustration}
    \end{center}
  \end{figure}

  Let us consider the time evolution of the BBS with box capacity $\Delta_n$
  from time $t$ to $t+1$.
  Assume that $Q^{(t+1)}_j$, $j=0, 1, \dots, n-1$, and
  $E^{(t+1)}_j$, $j=1, 2, \dots, n-1$, are given
  (see \Fref{fig:boxcap-ex-illustration}).
  Let $m$ be the index of the box which contains the leftmost segment of 
  the ($n+1$)th empty block at time $t$.
  Then the capacity of the $m$th box $\Delta_m$ is equal to
  $\Lambda^{(t)}_{n+1}$ by definition.
  Moreover, the relation 
  \begin{equation*}
    D^{(t+1)}_n
    =\sum_{j=0}^{n} Q^{(t)}_j-\sum_{j=0}^{n-1}Q^{(t+1)}_j
    =\sum_{j=-\infty}^{m-1}(U^{(t)}_j-U^{(t+1)}_j)+U^{(t)}_m,
  \end{equation*}
  where $U^{(t)}_k$ denotes the number of balls in the $k$th box at time $t$,
  also holds by definition.
  Hence, we can calculate the quantity $U^{(t+1)}_m$ by 
  the nu-KdV equation \eqref{eq:nu-KdV}:
  \begin{eqnarray*}
    U^{(t+1)}_m
    &=\min\left(\Delta_m-U^{(t)}_m, \sum_{j=-\infty}^{m-1}(U^{(t)}_j-U^{(t+1)}_j)\right)\\
    &=\min(\Lambda^{(t)}_{n+1}-U^{(t)}_m, D^{(t+1)}_n-U^{(t)}_m).
  \end{eqnarray*}
  Then we obtain the relation
  \begin{eqnarray*}
    \fl\Delta_m-U^{(t)}_m-U^{(t+1)}_m
    &=\Lambda^{(t)}_{n+1}-U^{(t)}_m-\min(\Lambda^{(t)}_{n+1}-U^{(t)}_m, D^{(t+1)}_n-U^{(t)}_m)\\
    &=-\min(0, D^{(t+1)}_n-\Lambda^{(t)}_{n+1})\\
    &=\max(0, \Lambda^{(t)}_{n+1}-D^{(t+1)}_n),
  \end{eqnarray*}
  where we have used the identity $-\min(-A, -B)=\max(A, B)$.
  This relation implies that the term $\max(0, \Lambda^{(t)}_{n+1}-D^{(t+1)}_n)$
  denotes the size of interspace inserted between the $n$th soliton
  at time $t$ and the $n$th one at time $t+1$.

  Once we notice the role of 
  the terms $\max(0, \Lambda^{(t)}_{n+1}-D^{(t+1)}_n)$,
  we can now clarify the meaning of the equations \eqref{eq:exToda-1} and \eqref{eq:exToda-2}.
  Since the term $E^{(t)}_{n+1}-\max(0, \Lambda^{(t)}_{n+1}-D^{(t+1)}_n)$ 
  in \eqref{eq:exToda-1}
  denotes the difference between the size of the inserted space and
  the one of the $(n+1)$th empty block,
  $Q^{(t+1)}_n$ should be determined by \eqref{eq:exToda-1}.
  Similarly, $E^{(t+1)}_n$ should be determined by \eqref{eq:exToda-2}.
  It is also true for $n=0$ and $1$,
  then the proof is completed by induction.
\end{proof}

\begin{figure}[htbp]
  \begin{center}
    \includegraphics[scale=1.15]{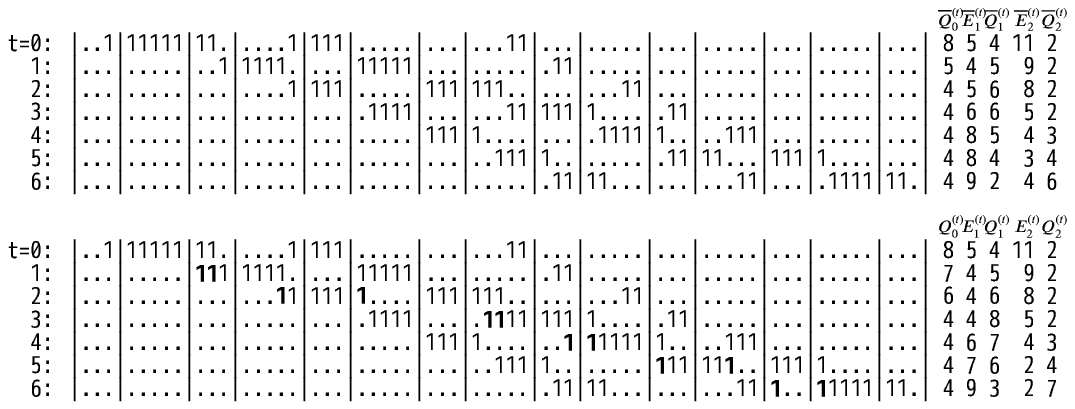}
  \caption{Example of the expansion map for the BBS with box capacity 
      $\Delta_n$ and carrier capacity $M_t=6$ for $t>0$. 
      This is obtained from the example 
      in Figure~\ref{fig:exampleOf2rnu-KP} via the expansion map.}
  \label{fig:exampleOfExpMapForGen}
  \end{center}
\end{figure}

Next, we construct the finite Toda representation of the BBS with both box
capacity and carrier capacity from two time evolution maps: 
the size limit map and the recovery map.
This is a similar as for the construction of the Euler representation 
explained in \sref{sec:euler-repr-bbs}.
\Fref{fig:exampleOfExpMapForGen} shows an example.

In the next theorem, we use the following notations:
\begin{itemize}
\item $Q^{(t)}_n$, $E^{(t)}_n$: 
  the size of the $n$th soliton and the one of the $n$th
  empty block at time $t$, respectively;
\item $\ol Q^{(t+1)}_n$, $\ol E^{(t+1)}_n$: the size of the $n$th soliton and 
  the one of the $n$th empty block after the size limit process from time 
  $t$ to $t+1$;
\item $\ol C^{(t+1)}_n$, $\ol D^{(t+1)}_n$: some quantities which will be
  explained in the proof of the next theorem in detail;
\item $K^{(t)}_n$, $\Lambda^{(t)}_n$: the capacity of the box which contains
  the leftmost segment of the $n$th soliton and the one of 
  the $n$th empty block at time $t$, 
  respectively.
\end{itemize}

\begin{theorem}\label{th:boxcap}
  Let the variables $Q^{(t)}_n$, $E^{(t)}_n$, 
  $\ol Q^{(t+1)}_n$, $\ol E^{(t+1)}_n$, $\ol C^{(t+1)}_n$ and $\ol D^{(t+1)}_n$ 
  denote the quantities of the BBS
  as explained in the above.
  Then the time evolution of the BBS with
  box capacity $K^{(t)}_n$ and $\Lambda^{(t)}_n$, and
  carrier capacity $M_{t+1}$ is given by
  \begin{subequations}\label{eq:enu-Toda}
    \begin{eqnarray}
      \ol Q^{(t+1)}_n=\min\left(E^{(t)}_{n+1}-\max(0, \Lambda^{(t)}_{n+1}-\ol D^{(t+1)}_n), \ol D^{(t+1)}_n\right),\label{eq:enu-Toda-olq}\\
      \eqalign{
      \ol E^{(t+1)}_n&=E^{(t)}_n-\ol Q^{(t+1)}_{n-1}+Q^{(t)}_n\\
      &\qquad-\max(0, \Lambda^{(t)}_n-\ol D^{(t+1)}_{n-1})+\max(0, \Lambda^{(t)}_{n+1}-\ol D^{(t+1)}_n),}\label{eq:enu-Toda-ole}\\
      \ol C^{(t+1)}_n=\min(\ol D^{(t+1)}_{n-1}-\ol Q^{(t+1)}_{n-1}+K^{(t)}_n, M_{t+1}),\label{eq:enu-Toda-olc}\\
      \ol D^{(t+1)}_n=\min(\ol C^{(t+1)}_n+Q^{(t)}_n-K^{(t)}_n, M_{t+1}),\label{eq:enu-Toda-old}\\
      Q^{(t+1)}_n=Q^{(t)}_n+\ol C^{(t+1)}_n-\ol C^{(t+1)}_{n+1}-K^{(t)}_n+K^{(t)}_{n+1},\label{eq:enu-Toda-q}\\
      E^{(t+1)}_n=\ol E^{(t+1)}_n+\ol Q^{(t+1)}_{n-1}-Q^{(t)}_n-\ol D^{(t+1)}_{n-1}+\ol D^{(t+1)}_n,\label{eq:enu-Toda-e}\\
      E^{(t)}_0=E^{(t)}_N=\ol E^{(t)}_0=\ol E^{(t)}_N=+\infty, \quad \ol C^{(t+1)}_0=K^{(t)}_0,\label{eq:enu-Toda-BC}
    \end{eqnarray}
  \end{subequations}
  where the carrier capacity $M_{t+1}$ must be chosen to
  satisfy the condition $K^{(t)}_n \le M_{t+1}$ for all $n$ and $t$.
\end{theorem}

When the quantities $\{Q^{(t)}_n\}_{n=0}^{N-1}$ and $\{E^{(t)}_n\}_{n=1}^{N-1}$
are given, first we can calculate $\ol D^{(t+1)}_0$ using 
\eqref{eq:enu-Toda-old} and \eqref{eq:enu-Toda-BC}.
Next, we can calculate $\ol Q^{(t+1)}_0$ by \eqref{eq:enu-Toda-olq},
$\ol C^{(t+1)}_1$ by \eqref{eq:enu-Toda-olc}, $\ol D^{(t+1)}_1$ by 
\eqref{eq:enu-Toda-old}. In a repetitive manner, we can obtain the quantities
$\{\ol Q^{(t+1)}_n\}_{n=0}^{N-1}$, $\{\ol C^{(t+1)}_n\}_{n=0}^{N}$ and 
$\{\ol D^{(t+1)}n\}_{n=0}^{N-1}$.
Finally, we can calculate the quantities $\{\ol E^{(t+1)}_n\}_{n=1}^{N-1}$,
$\{Q^{(t+1)}_n\}_{n=0}^{N-1}$ and $\{E^{(t+1)}_n\}_{n=1}^{N-1}$ by
\eqref{eq:enu-Toda-ole}, \eqref{eq:enu-Toda-q} and \eqref{eq:enu-Toda-e},
respectively.
Hence the time evolution is determined by \eqref{eq:enu-Toda}.

If $M_{t+1}=+\infty$, then \eqref{eq:enu-Toda-olc} and \eqref{eq:enu-Toda-old}
reduce to $\ol C^{(t+1)}_n=\ol D^{(t+1)}_{n-1}-\ol Q^{(t+1)}_{n-1}+K^{(t)}_n$
and $\ol D^{(t+1)}_n=\ol C^{(t+1)}_{n}+Q^{(t)}_n-K^{(t)}_n$, respectively.
Thus we have the equation
$\ol D^{(t+1)}_{n}=\ol D^{(t+1)}_{n-1}-\ol Q^{(t+1)}_{n-1}+Q^{(t)}_n$
and, substituting them into \eqref{eq:enu-Toda-q} and \eqref{eq:enu-Toda-e},
we obtain $Q^{(t+1)}_n=\ol Q^{(t+1)}_n$ and $E^{(t+1)}_n=\ol E^{(t+1)}_n$.
Hence, in this case, the ultradiscrete system \eqref{eq:enu-Toda}
reduces to the system \eqref{eq:exToda}.
We can therefore say that the system \eqref{eq:enu-Toda} is 
a generalization of the system \eqref{eq:exToda}.

\begin{proof}
  Let us show that the equations 
  \eqref{eq:enu-Toda-olq}--\eqref{eq:enu-Toda-old} describe the size
  limit process and \eqref{eq:enu-Toda-q}--\eqref{eq:enu-Toda-e} describe
  the recovery process.
  
  First, we consider the size limit process.
  The equations 
  \eqref{eq:enu-Toda-olq} and \eqref{eq:enu-Toda-ole} have
  the same forms as of \eqref{eq:exToda-1} and \eqref{eq:exToda-2}.
  Thus we shall investigate the variables $\ol C^{(t+1)}_n$ and
  $\ol D^{(t+1)}_n$ which are defined by 
  \eqref{eq:enu-Toda-olc} and \eqref{eq:enu-Toda-old}.
  Suppose that the carrier capacity is chosen as $K^{(t)}_n \le M_{t+1}<+\infty$,
  $\ol Q^{(t+1)}_j$, $j=0, 1, \dots, n-1$, and 
  $\ol E^{(t+1)}_j$, $j=1, 2, \dots, n-1$, are given,
  and the quantity $\ol D^{(t+1)}_{n-1}$ denotes the number of
  balls which the carrier holds after getting
  $Q^{(t)}_{n-1}$ balls from boxes and restricting the number of the holding balls
  to $M_{t+1}$ balls.
  Since the inequality $\ol Q^{(t+1)}_{n-1} \le \ol D^{(t+1)}_{n-1}$ holds 
  from \eqref{eq:enu-Toda-olq},
  it is sufficient to consider the following two cases:
  the case of which the carrier drops off all balls temporarily 
  ($\ol D^{(t+1)}_{n-1}-\ol Q^{(t+1)}_{n-1}=0$) and the case of which
  the carrier has balls just before getting $Q^{(t)}_n$ balls 
  ($\ol D^{(t+1)}_{n-1}-\ol Q^{(t+1)}_{n-1}>0$).
  \begin{enumerate}
  \item If $\ol D^{(t+1)}_{n-1}-\ol Q^{(t+1)}_{n-1}=0$, 
    then $\ol C^{(t+1)}_n=\min(\ol D^{(t+1)}_{n-1}-\ol Q^{(t+1)}_{n-1}+K^{(t)}_n, M_{t+1})=K^{(t)}_n$ holds 
    from the assumption.
    We should note that, in this case, 
    the number of balls which the carrier holds is zero temporarily 
    before getting $Q^{(t)}_n$ balls.
    Thus, from \eqref{eq:enu-Toda-old},
    we have $\ol D^{(t+1)}_{n}=\min(Q^{(t)}_{n}, M_{t+1})$, 
    which indicates that the quantity $\ol D^{(t+1)}_n$
    is again the number of balls which the carrier holds after
    getting $Q^{(t)}_n$ balls and restricting the number of the holding balls 
    to $M_{t+1}$ balls.

  \item The case of $\ol D^{(t+1)}_{n-1}-\ol Q^{(t+1)}_{n-1}>0$.
    Let $m$ be the index of the box which contains the leftmost segment
    of the $n$th soliton at time $t$.
    Under the assumption, in the terms of the variables of
    the Euler representation \eqref{eq:2rnuKp},
    $\ol D^{(t+1)}_{n-1}-\ol Q^{(t+1)}_{n-1}>0$ implies that
    $\ol U^{(t+1)}_m=\Delta_m-U^{(t)}_m$ should hold 
    (see \Fref{fig:exampleOfBoxCapCarCap}).
    \begin{figure}[htbp]
      \begin{center}
        \includegraphics[scale=.7]{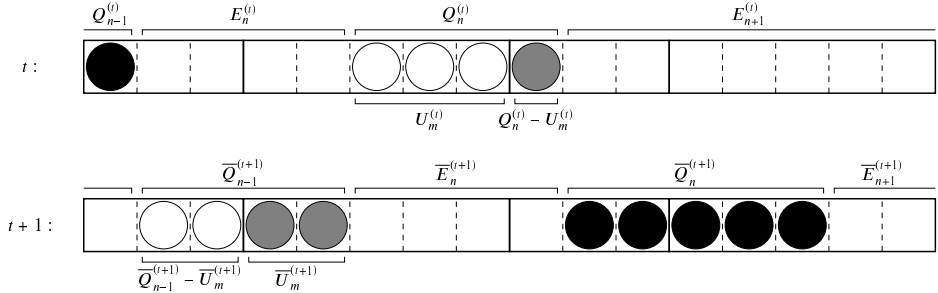}
        \caption{Illustration of the size limit process when
          the carrier parameter $M_{t+1}=6$ and 
          $\ol D^{(t+1)}_{n-1}-\ol Q^{(t+1)}_{n-1}>0$.
          We can see that $\ol C^{(t+1)}_n$ denotes the number of balls
          which the carrier holds after getting $U^{(t)}_m$ balls
          (white balls) and restricting the number of
          the holding balls to $M_{t+1}$,
          and $\ol D^{(t+1)}_n$ denotes the number of balls which the carrier holds
          after getting $Q^{(t)}_n-U^{(t)}_m$ balls (gray balls)
          and restricting the number of the holding balls to $M_{t+1}$ balls.} 
        \label{fig:exampleOfBoxCapCarCap}
      \end{center}
    \end{figure}
    \begin{table}[htbp]
      \caption{Change of the number of balls which the carrier holds.}
      \begin{center}
        \begin{tabular}{ll}
          State of the carrier & The number of balls which the carrier holds\\ \hline
          \vdots & \vdots\\ 
          Getting $Q^{(t)}_{n-1}$ balls & $\ol D^{(t+1)}_{n-1}$\\
          Putting $\ol Q^{(t+1)}_{n-1}-\ol U^{(t+1)}_m$ balls & $\ol D^{(t+1)}_{n-1}-(\ol Q^{(t+1)}_{n-1}-\ol U^{(t+1)}_m)$\\
          Getting $U^{(t)}_m$ balls & $\ol D^{(t+1)}_{n-1}-\ol Q^{(t+1)}_{n-1}+K^{(t)}_n$\\
          Size limit to $M_{t+1}$ balls & $\ol C^{(t+1)}_n$\\
          Putting $\ol U^{(t+1)}_m$ balls & $\ol C^{(t+1)}_n-\ol U^{(t+1)}_m$\\
          Getting $Q^{(t)}_n-U^{(t)}_m$ balls & $\ol C^{(t+1)}_n+Q^{(t)}_n-K^{(t)}_n$\\          
          Size limit to $M_{t+1}$ balls & $\ol D^{(t+1)}_n$\\
          \vdots & \vdots
        \end{tabular}
      \end{center}
      \label{tab:changeOfTheNumberOfBalls}
    \end{table}
    Now $\Delta_m=K^{(t)}_n$ by definition.
    Hence we can write \eqref{eq:enu-Toda-olc} and \eqref{eq:enu-Toda-old}
    as $\ol C^{(t+1)}_n=\min(\ol D^{(t+1)}_{n-1}-(\ol Q^{(t+1)}_{n-1}-\ol U^{(t+1)}_m)+U^{(t)}_m, M_{t+1})$
    and $\ol D^{(t+1)}_n=\min(\ol C^{(t+1)}_n+(Q^{(t)}_n-U^{(t)}_m)-\ol U^{(t+1)}_m, M_{t+1})$, respectively.
    Therefore, the quantity $\ol D^{(t+1)}_n$
    is again the number of balls which the carrier holds after
    getting $Q^{(t)}_n$ balls and restricting the number of the holding balls
    to $M_{t+1}$ balls.
    We can summarize the change of the number of balls which the carrier holds
    as \Tref{tab:changeOfTheNumberOfBalls}.
  \end{enumerate}
  Thus, together with the proof of Theorem \ref{th:boxcap},
  it is proved that \eqref{eq:enu-Toda-olq}--\eqref{eq:enu-Toda-old} 
  describe the size limit process by induction.

  Furthermore, since 
  the number of balls removed by the size limit process are given 
  by $(\ol D^{(t+1)}_{n-1}-\ol Q^{(t+1)}_{n-1}+K^{(t)}_n)-\ol C^{(t+1)}_n$
  and $(\ol C^{(t+1)}_n+Q^{(t)}_n-K^{(t)}_n)-\ol D^{(t+1)}_n$,
  we obtain the equations of the recovery process
  \begin{eqnarray*}
    \fl\eqalign{
      Q^{(t+1)}_n
      &=\ol Q^{(t+1)}_n+(\ol C^{(t+1)}_n+Q^{(t)}_n-K^{(t)}_n-\ol D^{(t+1)}_n)\\
      &\qquad\qquad+(\ol D^{(t+1)}_{n}-\ol Q^{(t+1)}_{n}+K^{(t)}_{n+1}-\ol C^{(t+1)}_{n+1})\\
      &=Q^{(t)}_n+\ol C^{(t+1)}_n-\ol C^{(t+1)}_{n+1}-K^{(t)}_n+K^{(t)}_{n+1},
    }\\
    \fl\eqalign{
      E^{(t+1)}_n
      &=\ol E^{(t+1)}_n-(\ol D^{(t+1)}_{n-1}-\ol Q^{(t+1)}_{n-1}+K^{(t)}_n-\ol C^{(t+1)}_n)\\
      &\qquad\qquad-(\ol C^{(t+1)}_n+Q^{(t)}_n-K^{(t)}_n-\ol D^{(t+1)}_n)\\
      &=\ol E^{(t+1)}_n+\ol Q^{(t+1)}_{n-1}-Q^{(t)}_n-\ol D^{(t+1)}_{n-1}+\ol D^{(t+1)}_n,
    }
  \end{eqnarray*}
  which lead to the equations \eqref{eq:enu-Toda-q} and \eqref{eq:enu-Toda-e},
  and the proof is completed.
\end{proof}

\begin{remark}
  Furthermore, the variables $X^{(t)}_0$ and $\ol X^{(t)}_0$,
  which denote the index of the leftmost segment of the 0th soliton 
  at time $t$ satisfy the equations 
  \begin{subequations}
    \begin{eqnarray*}
      \ol X^{(t+1)}_0=X^{(t)}_0+Q^{(t)}_0+\max(0, \Lambda^{(t)}_1-\ol D^{(t+1)}_0),\\
      X^{(t+1)}_0=\ol X^{(t+1)}_0-Q^{(t)}_0+\ol D^{(t+1)}_0=X^{(t)}_0+\max(\ol D^{(t+1)}_0, \Lambda^{(t)}_1).
    \end{eqnarray*}
  \end{subequations}
\end{remark}

\section{Particular solution for the fixed box capacity case}
\label{sec:part-solut-spec}

In this section, we discuss a particular solution
to the ultradiscrete system \eqref{eq:enu-Toda} 
with a special condition: all boxes have constant capacity $\Delta$.

Let us consider the bilinear equations
\begin{subequations}\label{eq:bilinears}
  \begin{eqnarray}
    \ol\tau^{0, t+1}_n\tau^{1, t-1}_n=\delta\tau^{0, t-1}_{n+1}\ol\tau^{1, t+1}_{n-1}+\tau^{0, t}_n\ol\tau^{1, t}_n,\label{eq:bilinear-olq}\\
    \delta\ol\tau^{0, t}_n\tau^{1, t}_n=(\delta-\mu_t)\tau^{0, t}_n\ol \tau^{1, t}_n+\mu_t\ol\tau^{0, t+1}_n\tau^{1, t-1}_n,\\
    \tau^{0, t}_{n+1}\ol\tau^{1, t}_n=\ol\tau^{0, t}_{n+1}\tau^{1, t}_n+\mu_{t}\ol\tau^{0, t+1}_n\tau^{1, t-1}_{n+1},\label{eq:bilinear-d}\\
    (\delta-\mu_t)\ol\tau^{0, t+1}_n\tau^{1, t-1}_{n+1}+\tau^{0, t}_{n+1}\ol\tau^{1, t}_n=\tau^{0, t-1}_{n+1}\ol\tau^{1, t+1}_n,\label{eq:bilinear-d2}
  \end{eqnarray}
\end{subequations}
where $\delta$ is a constant and $\mu_t$ is a parameter depending on $t$.
We introduce the dependent variables
\begin{eqnarray*}
  q^{(t)}_n=\frac{\ol\tau^{0, t+1}_{n+1}\tau^{1, t}_n}{\ol\tau^{0, t+1}_n\tau^{1, t}_{n+1}},\quad&
  \ol q^{(t)}_n=\frac{\delta}{\delta-\mu_t}\frac{\ol\tau^{0, t+1}_{n+1}\ol\tau^{1, t}_n}{\ol\tau^{0, t+1}_n\ol\tau^{1, t}_{n+1}},\\
  e^{(t)}_n=\delta^2\frac{\tau^{0, t}_{n+1}\ol\tau^{1, t+1}_{n-1}}{\tau^{0, t}_n\ol\tau^{1, t+1}_n},\quad&
  \ol e^{(t)}_n=\delta(\delta-\mu_t)\frac{\ol\tau^{0, t}_{n+1}\ol\tau^{1, t+1}_{n-1}}{\ol\tau^{0, t}_n\ol\tau^{1, t+1}_n},\\
  \ol c^{(t)}_n=\delta\frac{\ol\tau^{0, t}_n\tau^{1, t}_n}{\ol\tau^{0, t+1}_n\tau^{1, t-1}_n},\quad&
  \ol d^{(t)}_n=\frac{\delta}{\delta-\mu_t}\frac{\tau^{0, t}_{n+1}\ol\tau^{1, t}_n}{\ol\tau^{0, t+1}_n\tau^{1, t-1}_{n+1}}.
\end{eqnarray*}
Then \eqref{eq:bilinear-d2} yields the relation
\begin{equation*}
  1+\delta^{-1}d^{(t)}_n=(\delta-\mu_t)^{-1}\frac{\tau^{0, t-1}_{n+1}\ol\tau^{1, t+1}_n}{\ol\tau^{0, t+1}_n\tau^{1, t-1}_{n+1}}.
\end{equation*}
Further, \eqref{eq:bilinear-olq}--\eqref{eq:bilinear-d} yield the equations
\begin{subequations}\label{eq:end-Toda}
  \begin{eqnarray}
    \ol q^{(t+1)}_n=e^{(t)}_{n+1}(1+\delta^{-1}\ol d^{(t+1)}_n)+\ol d^{(t+1)}_n,\\
    \ol c^{(t+1)}_n=(\delta-\mu_{t+1})\frac{\ol d^{(t+1)}_{n-1}}{\ol q^{(t+1)}_{n-1}}+\mu_{t+1},\\
    \ol d^{(t+1)}_n=(\delta-\mu_{t+1})^{-1}\ol c^{(t+1)}_n q^{(t)}_n+\frac{\mu_{t+1}}{1-\delta^{-1}\mu_{t+1}},    
  \end{eqnarray}
  and the identities 
  \begin{eqnarray}
    \ol e^{(t+1)}_n=e^{(t)}_n\frac{q^{(t)}_n}{\ol q^{(t+1)}_{n-1}}\frac{1+\delta^{-1} \ol d^{(t+1)}_{n-1}}{1+\delta^{-1} \ol d^{(t+1)}_n},\\
    q^{(t+1)}_n=q^{(t)}_n\frac{\ol c^{(t+1)}_n}{\ol c^{(t+1)}_{n+1}},\\
    e^{(t+1)}_n=\ol e^{(t+1)}_n\frac{\ol q^{(t+1)}_{n-1} \ol d^{(t+1)}_n}{q^{(t)}_n \ol d^{(t+1)}_{n-1}},
  \end{eqnarray}
  hold. 
  In addition, we impose the finite lattice condition
  \begin{equation}
    e^{(t)}_0=e^{(t)}_N=\ol e^{(t)}_0=\ol e^{(t)}_N=0.
  \end{equation}
\end{subequations}
In the bilinear equations \eqref{eq:bilinears}, this condition implies
\begin{equation*}
  \tau^{k, t}_{-1}=\tau^{k, t}_{N+1}=\ol\tau^{k, t}_{-1}=\ol\tau^{k, t}_{N+1}=0.
\end{equation*}

We assume that the constant $\delta$ and the parameter $\mu_t$
satisfy the condition $0<\mu_t<\delta$ for all $t \in \mathbb Z$.
Then, putting $q^{(t)}_n=\rme^{-Q^{(t)}_n/\epsilon}$,
$e^{(t)}_n=\rme^{-E^{(t)}_n/\epsilon}$,
$\ol q^{(t)}_n=\rme^{-\ol Q^{(t)}_n/\epsilon}$,
$\ol e^{(t)}_n=\rme^{-\ol E^{(t)}_n/\epsilon}$,
$\ol c^{(t)}_n=\rme^{-\ol C^{(t)}_n/\epsilon}$,
$\ol d^{(t)}_n=\rme^{-\ol D^{(t)}_n/\epsilon}$,
$\delta=\rme^{-\Delta/\epsilon}$ into \eqref{eq:end-Toda}
and taking a limit $\epsilon \to +0$,
we obtain the ultradiscrete system \eqref{eq:enu-Toda} with 
the condition $K^{(t)}_n=\Lambda^{(t)}_n=\Delta \le M_{t+1}$
for all $n, t \in \mathbb Z$.

The following theorem is proved by using a determinant identity called
the Pl\"ucker relation.
\begin{theorem}
  A particular solution to the bilinear equations \eqref{eq:bilinears}
  with the semi-infinite lattice condition 
  $\tau^{k, t}_{-1}=\ol\tau^{k, t}_{-1}=0$ 
  \/ for all $k, t \in \mathbb Z$ is given by the Hankel determinants
  \begin{subequations}\label{eq:molecule solution}
    \begin{eqnarray}
      \tau^{k, t}_n=  
      \cases{
        0 & if $n<0$,\cr
        1 & if $n=0$,\cr
        |\xi^{(t)}_{k+i+j}|_{0 \le i, j \le n-1} & if $n>0$,}\\
      \ol\tau^{k, t}_n=    
      \cases{
        0 & if $n<0$,\cr
        1 & if $n=0$,\cr
        |\ol\xi^{(t)}_{k+i+j}|_{0 \le i, j \le n-1} & if $n>0$,}
    \end{eqnarray}
  \end{subequations}
  where $\xi^{(t)}_n$ and $\ol\xi^{(t)}_n$ are arbitrary functions
  satisfying the dispersion relation
  \begin{equation}\label{eq:dispersion}
    \ol\xi^{(t+1)}_n=-\delta \xi^{(t)}_{n+1}+\xi^{(t)}_n
    =(\mu_t-\delta)\ol\xi^{(t)}_{n+1}+\ol\xi^{(t)}_n,\quad n=0, 1, \dots.
  \end{equation}
\end{theorem}

Hereafter, we choose the arbitrary functions as
\begin{equation}\label{eq:xi eta}
  \fl \xi^{(t)}_n=\sum_{i=0}^{N-1} \frac{\eta^{(t)}_i}{p_i(p_i+\delta)^{n}},\quad
  \ol\xi^{(t)}_n=\sum_{i=0}^{N-1} \frac{\eta^{(t-1)}_i}{(p_i+\delta)^{n+1}},\quad
  \eta^{(t)}_i := \frac{w_i\prod_{j=0}^{t}(p_i+\mu_j)}{(p_i+\delta)^t},  
\end{equation}
where $p_i$ and $w_i$, $i=0, 1, \dots, N-1$, are some constants.
Then the dispersion relation \eqref{eq:dispersion} is satisfied and
the finite lattice condition 
$\tau^{k, t}_{-1}=\tau^{k, t}_{N+1}=\ol\tau^{k, t}_{-1}=\ol\tau^{k, t}_{N+1}=0$
holds for all $k, t \in \mathbb Z$.
Substituting \eqref{eq:xi eta} to \eqref{eq:molecule solution} and
expanding the Hankel determinants using the Cauchy-Binet formula, we obtain
\begin{eqnarray*}
  \fl\tau^{k, t}_n=\sum_{0 \le r_0<r_1<\dots<r_{n-1}\le N-1}\left(\left(\prod_{0\le i<j\le n-1}\frac{p_{r_i}-p_{r_j}}{(p_{r_i}+\delta)(p_{r_j}+\delta)}\right)^2 \prod_{i=0}^{n-1}\frac{w_{r_i}\prod_{j=0}^t(p_{r_i}+\mu_j)}{p_{r_i}(p_{r_i}+\delta)^{t+k}}\right),\\
  \fl\ol\tau^{k, t}_n=\sum_{0 \le r_0<r_1<\dots<r_{n-1}\le N-1}\left(\left(\prod_{0\le i<j\le n-1}\frac{p_{r_i}-p_{r_j}}{(p_{r_i}+\delta)(p_{r_j}+\delta)}\right)^2 \prod_{i=0}^{n-1}\frac{w_{r_i}\prod_{j=0}^{t-1}(p_{r_i}+\mu_j)}{(p_{r_i}+\delta)^{t+k}}\right),
\end{eqnarray*}
for $n=1, 2, \dots, N$.
These expressions can be ultradiscretized directly:
putting $p_n=\rme^{-P_n/\epsilon}$, 
$w_n=\rme^{-W_n/\epsilon}$,
$\tau^{k, t}_n=\rme^{-T^{k, t}_n/\epsilon}$,
$\ol\tau^{k, t}_n=\rme^{-\ol T^{k, t}_n/\epsilon}$,
and taking a limit $\epsilon \to +0$, 
we obtain the next theorem.

\begin{theorem}
  A particular solution to 
  the ultradiscrete system \eqref{eq:enu-Toda} with the condition
  $K^{(t)}_n=\Lambda^{(t)}_n=\Delta \le M_{t+1}$ for all $n, t \in \mathbb Z$
  is given by
  \begin{eqnarray*}
    \fl Q^{(t)}_n=\ol T^{0, t+1}_{n+1}-\ol T^{0, t+1}_n+T^{1, t}_n-T^{1, t}_{n+1},\quad&
    \ol Q^{(t)}_n=\ol T^{0, t+1}_{n+1}-\ol T^{0, t+1}_n+\ol T^{1, t}_n-\ol T^{1, t}_{n+1},\\
    \fl E^{(t)}_n=T^{0, t}_{n+1}-T^{0, t}_n+\ol T^{1, t+1}_{n-1}-\ol T^{1, t+1}_n+2\Delta,\quad&
    \ol E^{(t)}_n=\ol T^{0, t}_{n+1}-\ol T^{0, t}_n+\ol T^{1, t+1}_{n-1}-\ol T^{1, t+1}_n+2\Delta,\\
    \fl \ol C^{(t)}_n=\ol T^{0, t}_n-\ol T^{0, t+1}_n+T^{1, t}_n-T^{1, t-1}_n+\Delta,\quad&
    \ol D^{(t)}_n=T^{0, t}_{n+1}-\ol T^{0, t+1}_n+\ol T^{1, t}_n-T^{1, t-1}_{n+1},
  \end{eqnarray*}
  \begin{eqnarray*}
    \fl T^{k, t}_n=\min_{0\le r_0<r_1<\dots<r_{n-1}\le N-1}\Bigg(
      \sum_{i=0}^{n-1}\Big(W_{r_i}+\big(2(n-1-i)-1\big)P_{r_i}\\
      \fl\qquad\qquad-\big(2(n-1)+t+k\big)\min(P_{r_i}, \Delta)+\sum_{j=0}^t \min(P_{r_i}, M_j)\Big)\Bigg),\quad
      n=1, 2, \dots, N,\\
      \fl \ol T^{k, t}_n=\min_{0\le r_0<r_1<\dots<r_{n-1}\le N-1}\Bigg(
      \sum_{i=0}^{n-1}\Big(W_{r_i}+2(n-1-i)P_{r_i}\\
      \fl\qquad\qquad-\big(2(n-1)+t+k\big)\min(P_{r_i}, \Delta)+\sum_{j=0}^{t-1} \min(P_{r_i}, M_j)\Big)\Bigg),\quad
      n=1, 2, \dots, N,\\
      \fl T^{k, t}_{-1}=T^{k, t}_{N+1}=\ol T^{k, t}_{-1}=\ol T^{k, t}_{N+1}=+\infty,\quad
      T^{k, t}_0=\ol T^{k, t}_0=0,
  \end{eqnarray*}
  where $P_i$ and $W_i$, $i=0, 1, \dots, N-1$, are some constants satisfying
  $P_0 \le P_1 \le \dots \le P_{N-1}$.
\end{theorem}

\begin{remark}
  There exists a B\"acklund transformation from the discrete system 
  \eqref{eq:end-Toda} to the nonautonomous discrete Toda (nd-Toda) lattice:
  \begin{eqnarray*}
    \fl \mathsf q^{(t)}_n=\frac{\delta^{-1} q^{(t)}_n}{\delta(1+\delta^{-1}q^{(t)}_n)(1+\delta^{-1}e^{(t)}_n)},\quad&
    \mathsf e^{(t)}_n=\frac{\delta^{-1} e^{(t)}_n}{\delta(1+\delta^{-1}q^{(t)}_{n-1})(1+\delta^{-1}e^{(t)}_n)},\\
    \fl \ol{\mathsf q}^{(t)}_n=\frac{\delta^{-1} \ol q^{(t)}_n}{(\delta-\mu_t)(1+\delta^{-1}\ol q^{(t)}_n)(1+\delta^{-1}\ol e^{(t)}_n)},\quad&
    \ol{\mathsf e}^{(t)}_n=\frac{\delta^{-1} \ol e^{(t)}_n}{(\delta-\mu_t)(1+\delta^{-1}\ol q^{(t)}_{n-1})(1+\delta^{-1}\ol e^{(t)}_n)}.
  \end{eqnarray*}
  In fact, these variables have $\tau$-function expressions
  \begin{eqnarray*}
    \mathsf q^{(t)}_n=\delta^{-1}\frac{\tau^{0, t}_n\ol \tau^{0, t+1}_{n+1}}{\tau^{0, t}_{n+1}\ol \tau^{0, t+1}_n},\quad&
    \ol{\mathsf q}^{(t)}_n=(\delta-\mu_t)^{-1}\frac{\ol\tau^{0, t}_n\ol \tau^{0, t+1}_{n+1}}{\ol\tau^{0, t}_{n+1}\ol \tau^{0, t+1}_n},\\
    \mathsf e^{(t)}_n=\delta\frac{\tau^{0, t}_{n+1}\ol\tau^{0, t+1}_{n-1}}{\tau^{0, t}_n\ol\tau^{0, t+1}_n},\quad&
    \ol{\mathsf e}^{(t)}_n=(\delta-\mu_t)\frac{\ol\tau^{0, t}_{n+1}\ol\tau^{0, t+1}_{n-1}}{\ol\tau^{0, t}_n\ol\tau^{0, t+1}_n}.
  \end{eqnarray*}
  Since the bilinear equations
  \begin{eqnarray*}
    \tau^{0, t-1}_n\ol\tau^{0, t+1}_n=\delta(\delta-\mu_t)\tau^{0, t-1}_{n+1}\ol\tau^{0, t+1}_{n-1}+\tau^{0, t}_n\ol\tau^{0, t}_n,\\
    \delta\ol\tau^{0, t}_n\tau^{0, t}_{n+1}=(\delta-\mu_t)\tau^{0, t}_n\ol\tau^{0, t}_{n+1}+\mu_t\tau^{0, t-1}_{n+1}\ol\tau^{0, t+1}_n
  \end{eqnarray*}
  hold (these are proved by using the Pl\"ucker relation), 
  we have the equations
  \begin{equation*}
    \ol{\mathsf q}^{(t+1)}_n=\mathsf e^{(t)}_{n+1}+\ol{\mathsf d}^{(t+1)}_n,\quad
    \ol{\mathsf d}^{(t+1)}_n=\ol{\mathsf d}^{(t+1)}_{n-1}\frac{\mathsf q^{(t)}_n}{\ol{\mathsf q}^{(t+1)}_{n-1}}+\sigma_{t+1},
  \end{equation*}
  where
  \begin{eqnarray*}
    \ol{\mathsf d}^{(t)}_n:=(\delta-\mu_t)^{-1}\frac{\ol\tau^{0, t}_n\tau^{0, t}_{n+1}}{\tau^{0, t-1}_{n+1}\ol\tau^{0, t+1}_n},\quad
    \sigma_t:=\frac{\delta^{-1}\mu_{t}}{\delta-\mu_{t}}.    
  \end{eqnarray*}
  Additionally, we have the identities
  \begin{equation*}
    \fl \ol{\mathsf e}^{(t+1)}_n=\mathsf e^{(t)}_n\frac{\mathsf q^{(t)}_n}{\ol{\mathsf q}^{(t+1)}_{n-1}},\quad
    \mathsf q^{(t+1)}_n=\ol{\mathsf q}^{(t+1)}_n\frac{\ol{\mathsf d}^{(t+1)}_{n-1}\mathsf q^{(t)}_n}{\ol{\mathsf d}^{(t+1)}_n\ol{\mathsf q}^{(t+1)}_{n-1}},\quad
    \mathsf e^{(t+1)}_n=\ol{\mathsf e}^{(t+1)}_n\frac{\ol{\mathsf d}^{(t+1)}_n\ol{\mathsf q}^{(t+1)}_{n-1}}{\ol{\mathsf d}^{(t+1)}_{n-1}\mathsf q^{(t)}_n}.
  \end{equation*}
  Eliminating $\ol{\mathsf d}^{(t+1)}_n$ from these equations,
  we obtain the modified nd-Toda lattice \cite{maeda2010bbs}
  \begin{eqnarray*}
    \ol{\mathsf q}^{(t+1)}_n+\ol{\mathsf e}^{(t+1)}_n=\mathsf q^{(t)}_n+\mathsf e^{(t)}_{n+1}+\sigma_{t+1},\quad
    \ol{\mathsf q}^{(t+1)}_{n-1}\ol{\mathsf e}^{(t+1)}_n=\mathsf q^{(t)}_n\mathsf e^{(t)}_n,\\
    \mathsf q^{(t+1)}_n+\mathsf e^{(t+1)}_{n+1}=\ol{\mathsf q}^{(t+1)}_n+\ol{\mathsf e}^{(t+1)}_{n+1}-\sigma_{t+1},\quad
    \mathsf q^{(t+1)}_n\mathsf e^{(t+1)}_n=\ol{\mathsf q}^{(t)}_n\ol{\mathsf e}^{(t)}_n,
  \end{eqnarray*}
  and the finite lattice condition is given by
  \begin{equation*}
    \mathsf e^{(t)}_0=\mathsf e^{(t)}_N=\ol{\mathsf e}^{(t)}_0=\ol{\mathsf e}^{(t)}_N=0.
  \end{equation*}
\end{remark}

\section{Concluding remarks}\label{sec:concluding-remarks}
In this paper, we have derived the finite Toda representation of
the BBS with box capacity by introducing the expansion map from
a state of the BBS to a binary sequence.
Furthermore, we have given a particular solution for 
the fixed box capacity case.
Hence we can say that the ultradiscrete system \eqref{eq:enu-Toda} is
integrable if the parameters $K^{(t)}_n$ and $\Lambda^{(t)}_n$ are 
chosen as constants. 
Since there is a connection between the ultradiscrete system \eqref{eq:enu-Toda}
and the BBS with variable box capacity which is integrable,
we expect that the ultradiscrete system \eqref{eq:enu-Toda}
of the variable box capacity case is also integrable and
a discrete system derived through the inverse-ultradiscretization
has determinant solutions.
This problem is left for future research.

In the proof of Theorem \ref{th:boxcap}, 
the variables $\ol C^{(t+1)}_n$ and $\ol D^{(t+1)}_n$ have played
important roles; these variables denote the number of balls which 
the carrier has. 
Moreover, these variables correspond to the variables 
which are introduced to remove subtractions in the discrete equations
(see Remark \ref{rem:negative-problem}).
This result gives us a guideline for ultradiscretization of Toda-type
integrable systems and making connections between these systems and BBSs.

In 2000, Spiridonov and Zhedanov \cite{spiridonov2000stc} proposed
a Toda-type nonautonomous integrable system called {\em \Rii chain},
which is derived as the compatibility conditions of spectral transformations
for some biorthogonal rational functions.
By using techniques developed in this paper, we will be able to 
ultradiscretize the \Rii chain and consider a corresponding BBS.

\ack
The author thanks Professor Satoshi Tsujimoto
for fruitful discussions and helpful suggestions.
This work was supported by JSPS KAKENHI (11J04105).

\end{document}